# Optical whirlpool near absorbing metallic nanoparticle

M. V. Bashevoy,* V. A. Fedotov, and N. I. Zheludev
*EPSRC NanoPhotonics Portfolio Centre, School of Physics and Astronomy, University of Southampton, SO17 1BJ, United Kingdom*

The power-flow lines of light interacting with a metallic nanoparticle, in the proximity of its plasmon resonance, form whirlpool-like nanoscale optical vortices. Two different types of vortex have been detected. The outward vortex first penetrates the particle near its centerline then, on exiting the particle, the flow-lines turn away from the centerline and enter a spiral trajectory. Outward vortexes are seen for the wavelengths shorter then the plasmon resonance. For the wavelengths longer that the plasmon resonance the vortex is inward: the power-flow lines pass around the sides of the particle before turning towards the centerline and entering the particle to begin their spiral trajectory.

The structures of optical fields around metallic nanoparticles are of special interest due to their role in nanophotonic and plasmonic devices and meta-waveguides [1–3]. Here for the fist time, we report that light interacting with an absorbing metallic nanoparticle follows curly trajectories with curvatures on the sub-wavelength scale, creating whirlpool-like nanoscale optical vortices. These "energy sink" vortices with spiral energy flow line trajectories are seen in the proximity of the nanoparticle's plasmon resonance.

Optical vortices have been identified as features in scalar wavefront dislocations of monochromatic light fields and modal lines corresponding to non-monochromatic light as well as in singularities in the maps representing vectorial properties of light [4]. It is now recognized that singularities are often features of fields near sub-wavelength structures. A vortex structure in the streamlines of the Poynting vector has been detected for Sommerfeld's edge diffraction with discussion of the eel-like motion of light at the edge dating back to Newtonian times [5]. Recently vortices were found in light diffracted by narrow slits in silver and silicon [6, 7]. However, to the best of our knowledge, vortex field structures have never been detected in the vicinity of metal nanoparticles.

We studied the interaction of light with homogeneous isotropic spherical nanoparticles using Mie theory [8] — an exact analytical wave theory giving time-harmonic electromagnetic fields $\mathbf{E}$ and $\mathbf{H}$ at frequency $\omega$ that satisfy the wave equations

$$\bigtriangledown^2\mathbf{E} + k^2\mathbf{E} = 0, \quad \bigtriangledown^2\mathbf{H} + k^2\mathbf{H} = 0, \tag{1}$$

where $k^2 = \omega^2\varepsilon\mu$. Solutions to these equations are presented in the form of a series of spherical Bessel Functions inside the particle and spherical Hankel functions outside it. The nanoparticle is assumed to have a dielectric coefficient $\varepsilon$ and permittivity $\mu$. Mie theory gives exact solutions of the vector wave equation for the internal and scattered fields of the particle and has generated a massive body of literature in which field patterns for angle-dependant scattering, modes of excitation, and integral characteristics such as absorption and scattering cross-section have been calculated [9, 10]. It has been shown that light can bend near a nanoparticle [11], however it has never been determined that the interaction of light with a nanoparticle can create a nanoscale vortex field structure. Here we refer to vortices in the "trajectory" of light near the nanoparticle as defined by the lines of powerflow, i.e. lines to which the Poynting vector $\mathbf{P} = [\mathbf{E} \times \mathbf{H}]$ is tangential. In the vortex regime of propagation the lines of powerflow are wound around the nanoparticle to create a nanoscale "whirlpool", comparable in size to the particle itself, whereby light seems to pass through the particle several times over.

We found that the vortex regime occurs in metallic (e.g. silver) nanoparticles in the vicinity of the plasmon absorbtion resonance. We analyzed the field structure around a nanoparticle excited by a plane electromagnetic wave. To illustrate the vortex structures graphically, we plotted solutions in the plane of polarization of the incident light using powerflow lines and a color scale for the absolute value of the Poynting vector (reg = high, blue = low). In the field maps presented below the incident light is polarized in the plane of the page and propagates from left to right.

To relate the parameter field for our calculations to observable values we shall define the dimensionless scattering $\sigma_s$ and absorbtion $\sigma_a$ cross-sections of the nanoparticle. In the Rayleigh approximation, cross-sections for a particle much smaller than the wavelength are introduced via the particle's polarizability $\alpha$ and its geometrical cross-section $S$: $\sigma_s = (k^4/6\pi)|\alpha|^2/S$ and $\sigma_a = k\mathrm{Im}(\alpha)/S - \sigma_s$, where $k = 2\pi/\lambda$ is the wave vector, and polarizability is a function of the particle's shape and size [12].

We found that the existence of the vortex structure and the topology of the field maps depend on the values of the real and imaginary parts of the particle's complex dielectric coefficient $\varepsilon = \varepsilon' + i\varepsilon''$ (see Fig. 2). Here and below we assume non-magnetic nanoparticles with $\mu = 1$. Figures 2(a) and (b) show the modification of the field structure around a hypothetical nanoparticle for different values of $\varepsilon''$. In the case depicted in Fig. 2(a) the scattering and absorption cross-sections are much smaller than the geometrical cross-section and the particle is almost invisible to the external field ($\sigma_a = 0.47$, $\sigma_s = 0.03$). Most of the powerflow lines pass by the nanoparticle and only handful of them terminate on the particle, indicating small losses. In the case depicted in Fig. 2(b) the absorption cross-section approaches the plasmon resonance ($\sigma_a = 3.6$, $\sigma_s = 0.24$). Many flow-lines terminate at the nanoparticle (entering it from the front and the back), indicating high losses. When the absorbtion and scattering cross-sections increase even further the flow lines create vortex-like structures around the nanoparticle. Figures 2(c) and (d) show such vortices around a silver nanoparticle at wavelengths of 354 nm (where $\varepsilon = -2.0 + i0.28$, $\sigma_a = 5.8$ and $\sigma_s = 1.8$) and 367 nm (where $\varepsilon = -2.71 + i0.25$, $\sigma_a = 4.1$ and $\sigma_s = 2.0$).

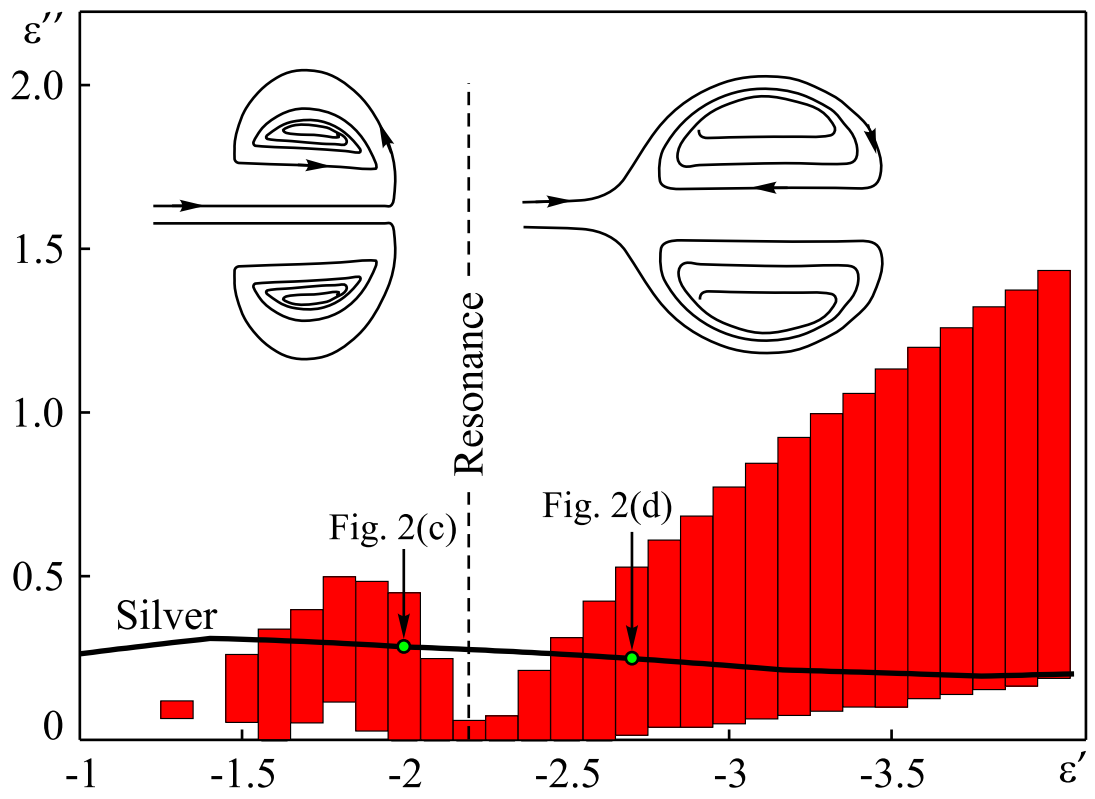

FIG. 1: Map showing values of the real and imaginary parts of the dielectric constant (in red) at which vortex field structures appear. The dashed line at $\varepsilon' \sim -2.2$ indicates the position of the plasmon resonance in a spherical nanoparticle with $r \approx 20$ nm ($\lambda/r = 20$). The solid lines show the dispersion characteristics of the dielectric properties of silver.

Figure 1 shows the parameter field where vortex structures can be observed. Two different types of vortex have been seen. In the first type, which we call an outward vortex, a bunch of powerflow lines first penetrate the particle near its centerline then, on exiting the particle, they separate, turn away from the centerline and enter a spiral trajectory. Outward vortices are seen to the "left" of the the plasmon resonance i.e. for $\varepsilon' > -2.2$ (in a spherical nanoparticle with a radius of 20 nm the plasmon resonance occurs at $\varepsilon' \sim -2.2$). In the second type of vortex, which we call an inward vortex, the powerflow lines pass around the sides of the particle before turning towards the centerline and entering the particle to begin their spiral trajectory. Outward vortexes are seen to the "right" of the the plasmon resonance i.e. for $\varepsilon' < -2.2$.

We also found by numerical simulation that vortex fields can exist near non-spherical nano-objects. Non-spherical nanoparticles are of considerable interest for applications because flattened or elongated shapes tend to reduce the plasmon resonance frequency, moving it from the blue-UV part of the spectrum to the more accessible visible-IR range. Mie theory is unsuitable for objects without spherical symmetry but computational methods provide an alternative to the analytical approaches and allow consideration of vortex fields around complex nanostructures. To analyze the vortex fields near spheroidal nanoparticles we used 64-bit software, developed by Comsol Inc., which implements a true 3D finite element method [14] and applies Perfectly Matched Layer (PML) [15] boundary conditions on all sides of the computational domain. We investigated a homogeneous oblate spheroidal nanoparticle with an aspect ratio of 2. Figure 3 shows the modification of the field structure around a spheroidal nanoparticle for different values of $\varepsilon''$. Here again, one can see the weak interaction regime in Fig. 3(a) ($\sigma_a = 0.42$, $\sigma_s = 0.02$), the high-loss regime in Fig. 3(b) ($\sigma_a = 3.7$, $\sigma_s = 0.3$), the creation of outward vortexes in Fig. 3(c) ($\sigma_a = 8.7$, $\sigma_s = 2.9$), and the creation of inward vortexes in Fig. 3(d) ($\sigma_a = 2.9$, $\sigma_s = 1.3$).

There are a number of intriguing questions that may be asked in relation to the nanoscale structuring of the energy flow near and inside the nanoparticle. For instance, a vortex structure with light passing through a nanoparticle several times backwards and forwards, resembles a standing wave in a dissipative Fabry-Perot resonator. One may therefore wonder if such a "nano-resonator" could provide conditions for a hysteresis and bistability in the nanoparticle's optical response if its dielectric properties depend on the intensity of light. The existence of vortex structures in nanoparticles could provide a graphical interpretation of the fact that the absorbtion cross-section of a particle can be much bigger that its geometrical cross-section. When a vortex is created, powerflow lines pass through the nanoparticle several times, "multiplying" the light-matter interaction and generating the high energy losses associated with the large optical cross-section.

The authors would like to thank M.V.Berry for important comments and useful references and K.F. MacDonald for discussions and assistance with manuscript preparation and also to acknowledge the support of the Engineering and Physical Sciences Research Council (UK).

* Corresponding author, e-mail: bmv@soton.ac.uk

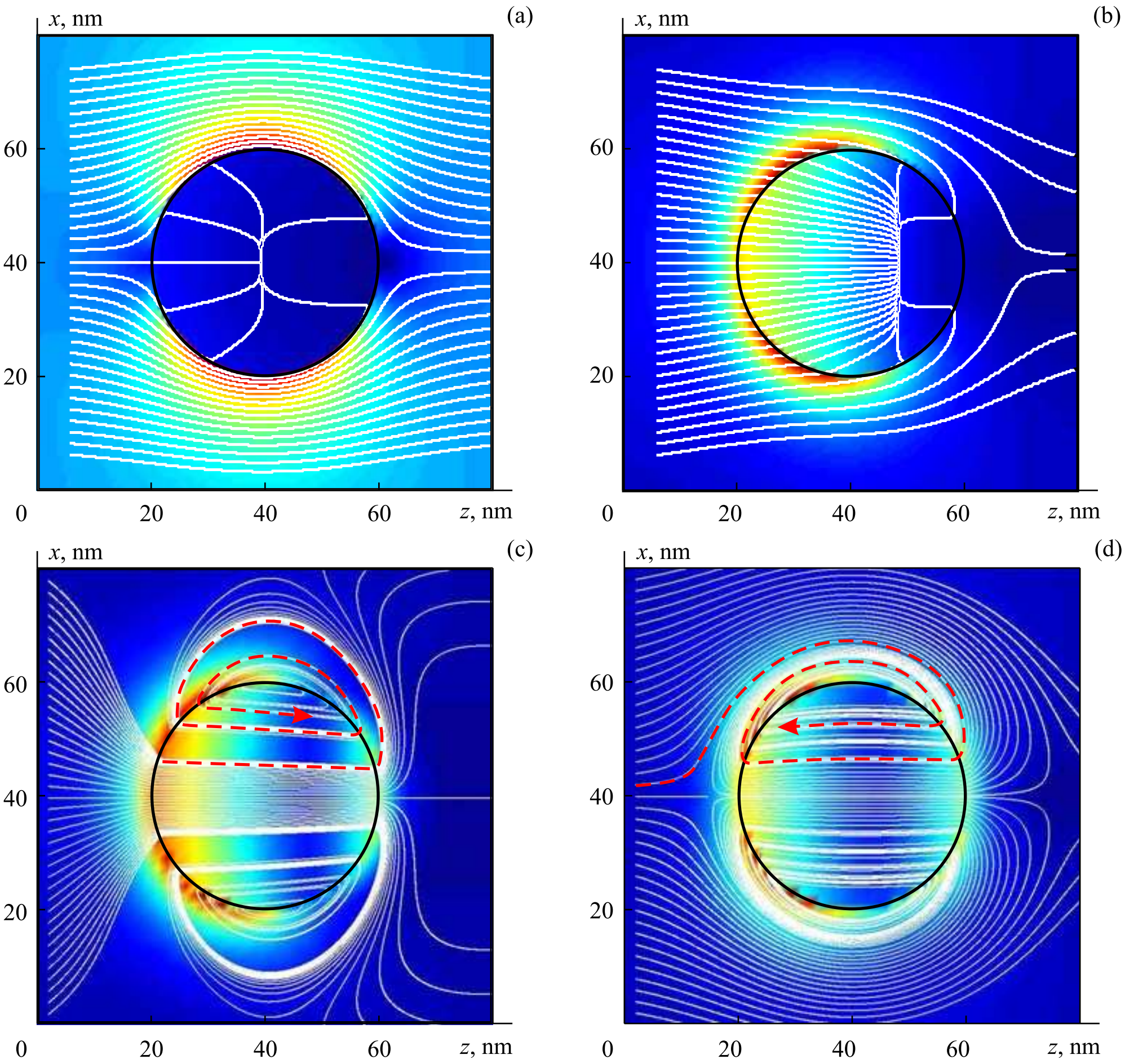

FIG. 2: Mie Theory: powerflow distribution around a spherical nanoparticle with a radius of approximately 20 nm ($\lambda/r = 20$) in the plane containing the directions of propagation (from left to right) and polarization of the incident light. The colors indicate the absolute value of the Poynting vector, the white lines show the direction of powerflow. (a) $\varepsilon = -2.0 + i10.0$, $\lambda = 400$ nm; (b) $\varepsilon = -2.0 + i1.0$, $\lambda = 400$ nm; (c) $\varepsilon = -2.0 + i0.28$ — the dielectric coefficient of silver at $\lambda = 354$ nm. Red dashed lines indicate outward vortex structure; (d) $\varepsilon = -2.71 + i0.25$ — the dielectric coefficient of silver at $\lambda = 367$ nm. Red dashed lines indicate inward vortex structure [13].

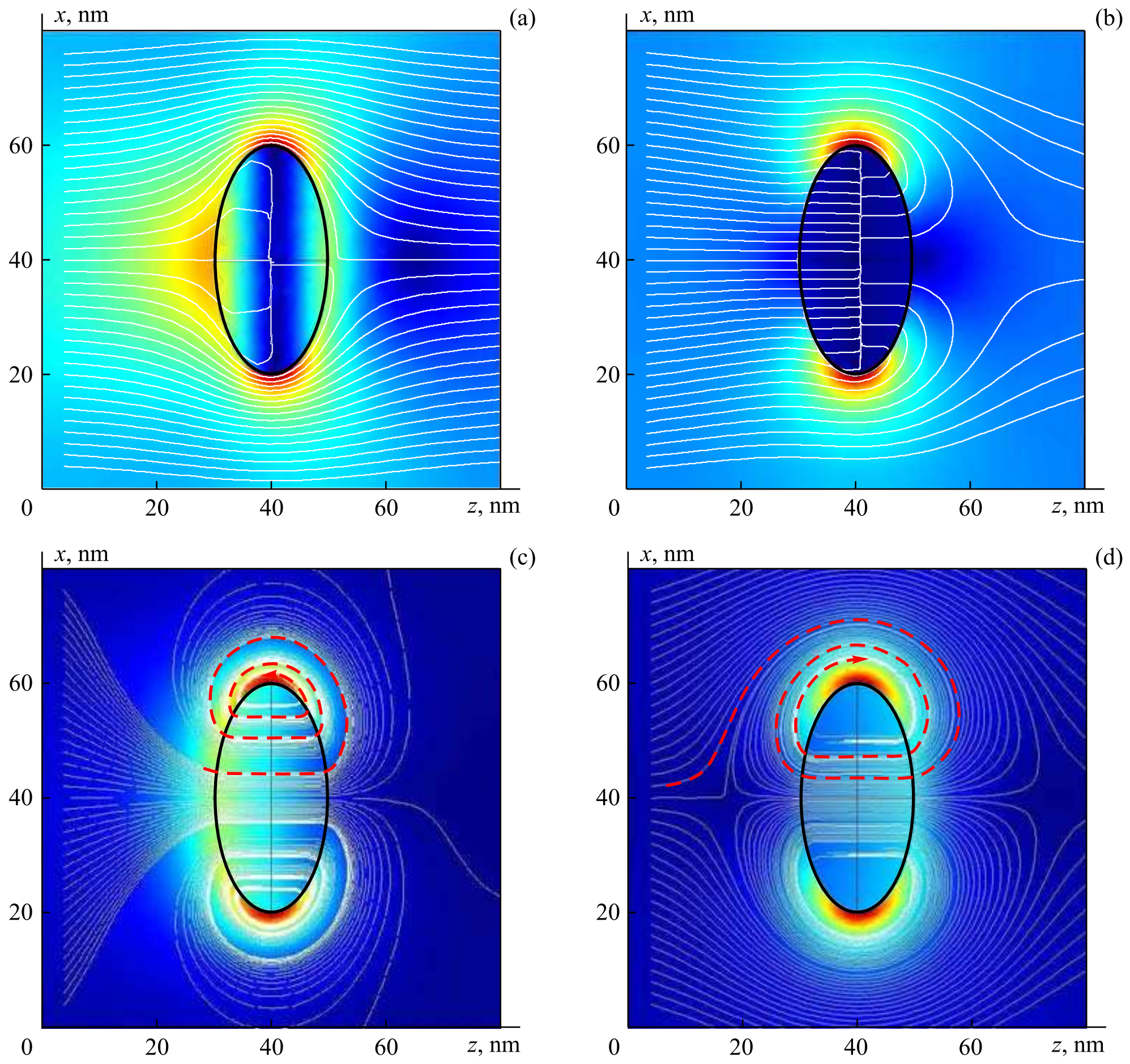

FIG. 3: 3D finite element modelling: powerflow distribution around an oblate spheroidal nanoparticle (with a semi-major axial radius of approximately 20 nm ($\lambda/r = 20$) and an aspect ratio of 2) in the plane containing the directions of propagation (from left to right) and polarization of the incident light. The colors indicate the absolute value of the Poynting vector, the white lines show the direction of powerflow. (a) $\varepsilon = -3.52 + i10.0$, $\lambda = 400$ nm; (b) $\varepsilon = -3.52 + i1.0$, $\lambda = 400$ nm; (c) $\varepsilon = -3.37 + i0.2$ — the dielectric coefficient of silver at $\lambda = 380$ nm. Red dashed lines indicate outward vortex structure; (d) $\varepsilon = -4.0 + i0.2$ — the dielectric coefficient of silver at $\lambda = 392$ nm. Red dashed lines indicate inward vortex structure [13].